\documentclass[sigconf,nonacm]{acmart}

\AtBeginDocument{%
  \providecommand\BibTeX{{%
    \normalfont B\kern-0.5em{\scshape i\kern-0.25em b}\kern-0.8em\TeX}}}

\usepackage{amsmath,amssymb,amsfonts}
\usepackage{textcomp} 
\usepackage{multicol} 
\usepackage[utf8]{inputenc}
\usepackage{listings}
\usepackage[xindy,style=index,numberedsection=false]{glossaries}
\usepackage{url}
\usepackage{multirow}
\usepackage[]{algpseudocode,algorithm}
\algtext*{EndWhile}
\algtext*{EndIf}
\algtext*{EndFor}
\algtext*{EndFunction}
\usepackage{textcomp}
\usepackage{cleveref}
\usepackage{booktabs}
\usepackage{makecell}
\usepackage{caption}
\usepackage{siunitx}

\usepackage{setspace}
\let\Algorithm\algorithm
\renewcommand\algorithm[1][]{\Algorithm[#1]\setstretch{1.0}}
\algnewcommand{\LineComment}[1]{\State \(\triangleright\) #1}

\crefname{algocf}{alg.}{alg.}
\Crefname{algocf}{Alg.}{Alg.}
\crefname{lstlisting}{listing}{listings}
\Crefname{lstlisting}{Listing}{Listings}
\crefname{algorithm}{alg.}{alg.}
\Crefname{algorithm}{Alg.}{Alg.}

\AtBeginDocument{\let\latexlabel\label} 

\newcommand{\ra}[1]{\renewcommand{\arraystretch}{#1}}
\newcommand{\pluseq}{\mathrel{+}=}

\newacronym{nlp}{NLP}{Natural Language Processing}
\newacronym{ppr}{PPR}{Personalized PageRank}
\newacronym{fpga}{FPGA}{Field-Programmable Gate Array}
\newacronym{spmv}{SpMV}{Sparse matrix-vector multiplication}
\newacronym{coo}{COO}{Coordinate}
\newacronym{dsl}{DSL}{Domain-Specific Language}
\newacronym{csc}{CSC}{Compressed Sparse Column}
\newacronym{raw}{RAW}{Read-After-Write}
\newacronym{ndcg}{NDCG}{Normalized Discounted Cumulative Gain}
\newacronym{ir}{IR}{Information Retrieval}
\newacronym{dcg}{DCG}{Discounted Cumulative Gain}

\begin{document}

\title[A reduced-precision streaming SpMV...]{A reduced-precision streaming SpMV architecture for Personalized PageRank on FPGA}

\author{Alberto Parravicini}
\email{alberto.parravicini@polimi.it}
\affiliation{%
  \institution{Politecnico di Milano}
  \city{Milan}
  \state{Italy}
}

\author{Francesco Sgherzi}
\email{francesco1.sgherzi@mail.polimi.it}
\affiliation{%
  \institution{Politecnico di Milano}
  \city{Milan}
  \state{Italy}
}

\author{Marco D. Santambrogio}
\email{marco.santambrogio@polimi.it}
\affiliation{%
  \institution{Politecnico di Milano}
  \city{Milan}
  \state{Italy}
}

\renewcommand{\shortauthors}{Parravicini et al.}

\begin{abstract}
Sparse matrix-vector multiplication is often employed in many data-analytic workloads in which low latency and high throughput are more valuable than exact numerical convergence. FPGAs provide quick execution times while offering precise control over the accuracy of the results thanks to reduced-precision fixed-point arithmetic. In this work, we propose a novel streaming implementation of Coordinate Format (COO) sparse matrix-vector multiplication, and study its effectiveness when applied to the Personalized PageRank algorithm, a common building block of recommender systems in e-commerce websites and social networks. Our implementation achieves speedups up to 6x over a reference floating-point FPGA architecture and a state-of-the-art multi-threaded CPU implementation on 8 different data-sets, while preserving the numerical fidelity of the results and reaching up to 42x higher energy efficiency compared to the CPU implementation.
\end{abstract}



\maketitle

\glsresetall
\section{Introduction}\label{sec:intro}

\gls{spmv} is a computational element widely employed in machine learning, engineering, and most importantly, graph analytics \cite{zhang2018shufflenet, kepner2016mathematical} as real-world graphs present an extremely high degree of sparsity.
\gls{ppr} \cite{bahmani2010fast}, a variation of the famous PageRank algorithm ranks the most relevant vertices of the graph with respect to an input vertex. In most cases \gls{ppr} must be computed with minimal latency, often on graphs with millions of edges, such as domain-specific knowledge bases, e-commerce websites, and social networks communities \cite{li2010computational, snapnets}, to find recommended posts in a social network while users interact with it, or recommended items for a given query on an e-commerce platform. 
Moreover, the precise numerical values produced by the algorithm are rarely useful, as long as the order of the top-ranked vertices is correct (consider the problem of recommending the top-10 products for a user query). 
Numerical boundedness of \gls{ppr} makes \glspl{fpga} suitable for computing \gls{ppr} with throughput beyond traditional architectures, leveraging fixed-point arithmetic that can reduce execution time while preserving the correct ranking, and accelerate convergence. 

In this work, we propose a novel \gls{fpga} architecture for a streaming edge-centric \gls{spmv} that uses \gls{coo} format matrices, and apply it to the computation of \gls{ppr}.
Reduced-precision fixed-point arithmetic is used to maximize performance while reducing resource utilization and preserving the quality of the results. 


In summary, we present the following contributions:
\begin{itemize}
    \item An optimized \gls{fpga} architecture of \gls{spmv} that leverages a \gls{coo} matrix and reduced-precision arithmetic, which we employ in a novel implementation of \gls{ppr} (\Cref{sec:implementation}).
    \item We validate the practical applicability of our \gls{ppr} implementation on 8 different graphs against a state-of-the-art multi-threaded CPU implementation and an equivalent  32-bits floating-point FPGA architecture, reaching speedups up to \textbf{6.8x} and up to \textbf{42x} higher energy efficiency.
    \item
    Most importantly, we characterize how reduced precision leads to negligible accuracy loss and \textbf{2x} faster convergence on \gls{ppr}, showing the effectiveness of reduced precision for approximate graph ranking algorithms
    (\Cref{sec:experimental_results}). 
\end{itemize}
\section{Related Work}

In this section, we provide an overview of existing research on the optimization of \gls{spmv} for different hardware architectures, especially in the context of graph algorithms and \gls{ppr}.  



\subsection{CPU and GPU Implementations}

Leveraging sparse linear algebra for graph processing is the focus of the GraphBLAS project, which aims at defining operations on graphs through the language of linear algebra \cite{kepner2016mathematical}, and it offers early implementations for both CPU and GPU \cite{bulucc2011combinatorial, yang2019graphblast}.
Highly tuned implementations of \gls{ppr} exploit the graph data-layout to maximize cache usage \cite{zhou2017design}, or employ multi-machine setups to process trillions of edges \cite{zhu2016gemini}.
Green-Marl \cite{hong2012green} and GraphIt \cite{zhang2018graphit} implements \gls{ppr} using \glspl{dsl} that abstract the intricacies of graph processing, and optimized to fully exploits the CPU hardware. 
\gls{ppr} on GPU is less common: it is worth mentioning nvGRAPH \cite{nvgraph} and GraphBLAST \cite{yang2019graphblast}, that leverage sparse linear algebra to match and possibly outperform CPU implementations.

\subsection{FPGA Implementations}

To the best of our knowledge, no existing work specifically addresses the computation of \gls{ppr} on \gls{fpga}, either using reduced-precision arithmetic or algorithmic optimizations.


However, there have been significant contributions in optimizing \gls{spmv} computations on \glspl{fpga}, as \gls{spmv} represents the main bottleneck of many PageRank implementations. Recent work by Grigoras et al. \cite{grigoras2015accelerating} focuses on compressing the sparse matrix, moving the bottleneck from memory accesses to the decompression of the input data while lowering the storage demand. Umuroglu et al. \cite{umuroglu2015vector} leverage local cache hierarchies and pre-processing schemes to maximize the amount of time in which values are kept in a fast local cache. Using data-set partitioning and complex memory hierarchies enable \gls{spmv} computations on web-scale graphs, as seen in Shan et al. \cite{shan2010fpga}: clearly, there is a performance trade-off introduced by supporting larger graphs, and simpler design might be more beneficial for smaller data-sets such as the ones in our intended use-case.
Reduced-precision arithmetic has not been thoroughly studied in the context of graph ranking algorithms, but encouraging results were shown in numerical analysis and deep-learning \cite{liang2018fp, wang2019deep}.
\section{Problem Definition}

\begin{figure}
    \centering
    \includegraphics[width=0.45\textwidth]{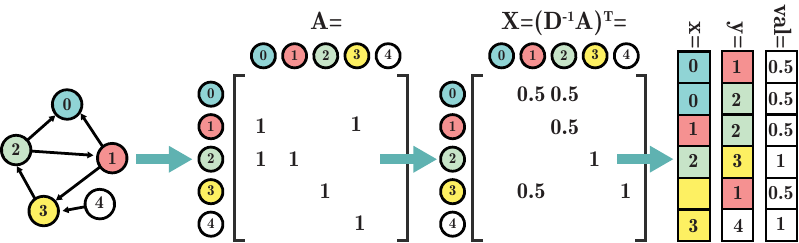}
    \caption{A graph as \gls{coo} matrix.
    In $\mathbf{X}$, each value $val$ can be seen as the probability of moving from $y$ to $x$. For example, from vertex $x=0$ there is a $0.5$ probability of coming from vertex $y=1$ and a $0.5$ probability of coming from vertex $y=2$}
    \label{fig:coo}
\end{figure}


In this work, we apply a novel \gls{spmv} architecture to the computation of \acrlong{ppr},  an algorithm that provides a \textit{personalized} ranking of the graph vertices, such that vertices that are more \textit{relevant} to an input vertex will have a higher score.

Given a graph $G$ with $|V|$ vertices and $|E|$ edges, we represent it using the adjacency matrix $\mathbf{A}$ and out-degree matrix $\mathbf{D}$ (a diagonal matrix with the number of out-going edges of each vertex). Define $\mathbf{X} = (\mathbf{D}^{-1}\mathbf{A})^T$ as the probability of transitioning from a vertex to one of its neighbors\footnote{assuming uniform probability, the probability of moving from a vertex $x$ with out-degree $d$ to a neighbor $y$ is $1/d$}, a \textit{personalization vertex} $v$, and a vector $\mathbf{p_t}$ of PageRank values, personalized w.r.t. $v$, computed at iteration $t$. $1 - \alpha$ is the probability of moving to any random vertex, and $\mathbf{\bar{d}}$ is a \textit{dangling} vector s.t. $\bar{d_i} = 1 \Leftrightarrow D_{ii} = 0,\ \bar{d_i} = 0 \Leftrightarrow D_{ii} \neq 0$. $\mathbf{\bar{d}}$ is added to $\mathbf{D}$ to ensure that the computation is numerically stable  \cite{ipsen2008pagerank}.
The vector $\mathbf{\bar{v}}$ is equal to $0$ except for the element at index $v$, which is $1$. The recurrence equation \cite[Section~3]{boldi2009pagerank} of \gls{ppr} is

\begin{equation}\label{eq:ppr}
    \mathbf{p_{t+1}} = \alpha\mathbf{X}\mathbf{p_t} + \frac{\alpha}{|V|}(\mathbf{\bar{d}}\mathbf{p_t})\mathbf{1} + (1-\alpha)\mathbf{\bar{v}}
\end{equation}

The first term of the right-hand side is a matrix-vector multiplication, while the second and third terms (the \textit{dangling factor} and the \textit{personalization factor}) are obtained with dot-products. The weighted adjacency matrix $\mathbf{X}$ is stored in a \textit{sparse format} as it is extremely sparse: in a graph with $10^6$ vertices and average out-degree $10$, only $10 \cdot 10^6 / 10^{12}$ (i.e. $0.001\%$) of the entries of $\mathbf{X}$ are non-zero. 

\gls{csc}, a common storage format for sparse matrices \cite{shan2010fpga}, can be inefficient for real-world graphs with vertex degrees that follow an exponential distribution, as it limits pipelined architectures that demand precise knowledge of data boundaries.
Instead, we employ the \gls{coo} storage layout (\cref{fig:coo}), which uses three equally sized arrays, containing, for each entry, its value and its two coordinates.
\gls{coo} simplifies array partitioning, enables burst reads from memory, and pipelined hardware designs, as entries are independent and the architecture is not bound to knowing the degree of each vertex. 
Instead, \gls{csc}-based designs often fail to handle graphs with exponential distribution, especially if stream-like processing is demanded.

We compute $\kappa$ \textit{personalization} vertices in parallel, to batch multiple  user requests. We replace $\mathbf{p_t}$ with a matrix $\mathbf{P_t}$ of size $|V| \times \kappa$, and $\mathbf{\bar{v}}$ with a matrix $\mathbf{\bar{V}}$.
Updating $\mathbf{P_t}$ requires reading all the edges only once. This optimization boosts the efficiency of a memory-bound algorithm, and enables higher throughput and scalability.

\section{Implementation}\label{sec:implementation}

We present the building blocks of our \gls{spmv} architecture and how we  integrated it in the \gls{ppr} computation, our intended use-case.

\subsection{Personalized PageRank Implementation}\label{sec:ppr_impl}

\begin{figure}
    \centering
    \includegraphics[width=0.47\textwidth]{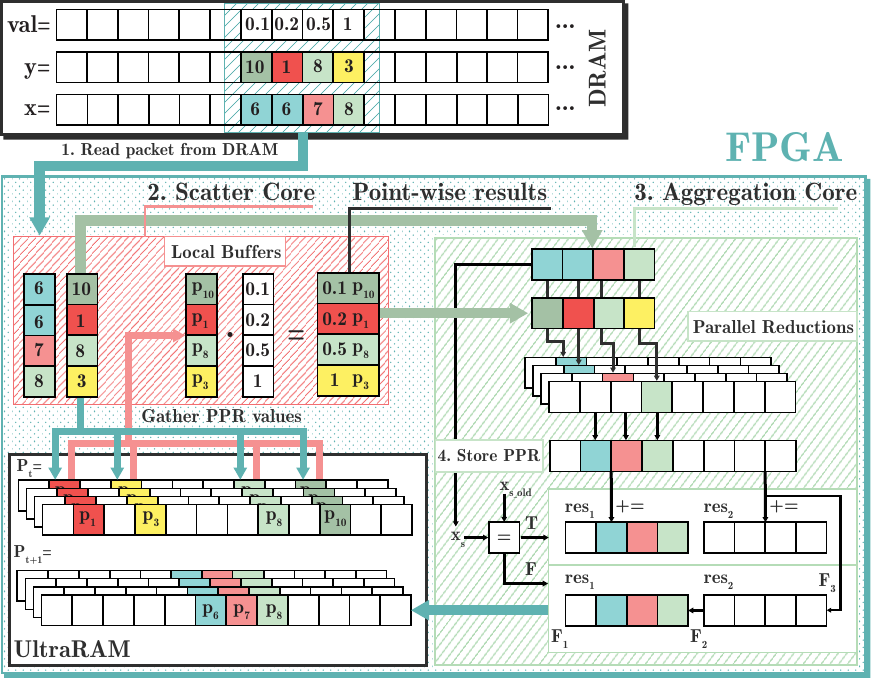}
    \caption{Representation of our \gls{spmv} architecture. The \textit{scatter} and \textit{aggregation} cores show the computation for a single vertex, but they are replicated to support $\kappa$ vertices. Large arrows represent a streaming transfer between cores}
    \label{fig:ppr}
\end{figure}

\Cref{alg:ppr} contains the pseudo-code of the main \gls{ppr} computation. The input graph is read from DRAM, with edges as packets of size $P\_SIZE=256$ to maximize the throughput of memory transactions, and process $B$ edges per clock cycle (8, if $P\_SIZE=256$ bits and each value is $32$ bits). Lines 6-8 of \Cref{alg:ppr} are the core of \gls{ppr}, with the \gls{spmv} computation further detailed in \cref{alg:spmv} and \cref{fig:ppr}. The $\kappa$ entries of the scaling vector are computed as the sum of current \gls{ppr} values of vertices with no outgoing edges. Values in the \textit{dangling} bitmap are read in blocks with size $P\_SIZE$ 
, while $\mathbf{P}$ is cyclically partitioned to access $B$ contiguous values in a single clock cycle.
\gls{ppr} values are stored as reduced-precision fixed-point values. Quantization truncates to zero the fractional bits with precision higher than representable. Other policies (e.g. rounding to the closest representable value) resulted in numerical instability.

\subsubsection{SpMV Design}\label{sec:spmv}

Our \gls{spmv} architecture has 4 main steps. First, we read a graph packet from DRAM (lines 4-5 in \cref{alg:spmv}), and store it in local buffers $x,\ y,\ val$ to read and update $B$ values at once. While we compute $\kappa$ \gls{ppr} vectors in parallel, the edges of the graph are accessed only once.
Parallel accesses to $\mathbf{P_t}$ retrieve \gls{ppr} values for each \textit{personalization} vertex: thanks to UltraRAM, we perform these accesses with low latency, without imposing strong constraints on the graph size.
The $B$ \textit{aggregator cores} (lines 12-17) combine \textit{point-wise} contributions to obtain the total contribution of a single vertex, as a packet can contain multiple edges referring to it. Each \textit{aggregator} considers edges whose end is in the range $[x[0], x[0] + B]$, i.e. the maximum range that can be found in a packet.

\begin{algorithm}[h]
  \caption{Personalized PageRank}\label{alg:ppr}
  \begin{algorithmic}[1]
    \Function{PPR}{$coo\_graph, \mathbf{\bar{V}}, \mathbf{\bar{d}}, \alpha, max\_iter$}
      \State Initialize local buffers to $0$
      \For{$k \gets 0,\kappa$} \Comment{Set PR=1 on pers. vertices}
        \State $\mathbf{P_1}[k] = \mathbf{\bar{V}}[k]$
      \EndFor
      \For{$i \gets 0,max\_iter$} 
        \State $scaling\_vec \gets scaling(\mathbf{P_1},\mathbf{\bar{d}})$ \Comment{$i.e.\ \frac{\alpha}{|V|}\mathbf{P_1}\mathbf{\bar{d}}$}
        
        \State $SpMV(coo\_graph, \mathbf{P_1}, \mathbf{P_2})$ \Comment{$\mathbf{Xp_i}$ in \cref{eq:ppr}}
        \State $\mathbf{P_1} = \alpha \mathbf{P_2} + scaling\_vec + (1-\alpha)\mathbf{\bar{V}}$
        
      \EndFor
      \State Write $\mathbf{P_1}$ to output
    \EndFunction
  \end{algorithmic}
\end{algorithm}

The last step adds \gls{ppr} contributions of the current packet to the \gls{ppr} arrays stored in UltraRAM.
Contributions are stored in a buffer of size $2B$, with up to $B$ non-zero contiguous values.
A Finite-State Machine with 2 buffers of size $B$ accumulates \gls{ppr} entries and writes them to output at indices multiple of $B$, ensuring that updates can be performed in parallel as they are aligned to the partitioning factor of $\mathbf{P_{t+1}}$.
Each block of $res_1$ is written on UltraRAM only once to avoid expensive $\pluseq$ operations and \gls{raw} conflicts in unrolled loops.
The 4 main steps of the algorithm, presented here as a single loop (\cref{alg:spmv}, line 2), are implemented as separate modules in a streaming \textit{data-flow} region, enabling aggressive pipelining of loops and better resource allocation.

\begin{algorithm}
  \caption{\gls{coo} \gls{spmv}}\label{alg:spmv}
  \begin{algorithmic}[1]
    \Function{SpMV}{$coo\_graph, \mathbf{P_t}, \mathbf{P_{t+1}}$}
      \For{$i \gets 0..|E| / B$}
        \LineComment{\textbf{1. Process \gls{coo} in packets of size $B$}}
        \State $x \gets coo\_graph.x[i]; y \gets coo\_graph.y[i]$
        \State $val \gets coo\_graph.val[i]$
        \State 
        
        \For{$k \gets 0..\kappa$} \Comment{$\kappa$ \textit{personalization vertices}}
            \LineComment{\textbf{2. Update edge-wise \gls{ppr} values}}
            \For{$j \gets 0..B$} 
                \State $dp\_buffer[k, j] = val[j] \cdot \mathbf{P_t}[k, y[j]]$
            \EndFor
            
            \LineComment{\textbf{3. Aggregate partial \gls{ppr} values}}
            \For{$b1 \gets 0..B$}
               \For{$b2 \gets 0..B$}
                    \State $agg\_res[k, x[0]\ \%\ B + b1] \pluseq$\par
                    \hskip\algorithmicindent $dp\_buffer[k, b2] \cdot ((x[0] + b1) == x[b2])$
                \EndFor
            \EndFor
            
            \LineComment{\textbf{4. Store \gls{ppr} values on each vertex}}
            \State $x_s \gets \lfloor x[0] / B\rfloor \cdot B$
            \If{$x_s == x_{s\_old}$}
                \For{$j \gets 0..B$} 
                    \State $res_1[k, j] \pluseq agg\_res[k, j]$
                    \State $res_2[k, j] \pluseq agg\_res[k, j + B]$
                \EndFor  
            \Else
                \For{$j \gets 0..B$} 
                    \State $res[k, j + x_{s_old}] = res_1[k, j]$
                    \State $res_1[k, j] = res_2[k, j] + agg\_res[k, j]$
                    \State $res_2[k, j] = agg\_res[k, j + B]$
                \EndFor
            \EndIf
            \State $reset(agg\_res); \ x_{s\_old} \gets x_s$
        \EndFor
      \EndFor
    \EndFunction
  \end{algorithmic}
\end{algorithm}

\subsubsection{PPR Buffers Design}\label{sec:uram}

Temporary \gls{ppr} values are stored in UltraRAM (URAM), a type of memory available in recent Xilinx UltraScale+\textsuperscript{TM} \glspl{fpga}. UltraRAM can be seen as a middle-ground between slow but abundant DRAM and faster, but limited, BRAM. Using a Xilinx Alveo U200 Accelerator Card, we store up to 90MB of data on UltraRAM, corresponding to around 20 million different PageRank values, assuming that the PageRank value of each vertex is stored in 32-bits. In practice, reduced fixed-point precision allows us to store even more vertices, and scale to larger graphs.
The maximum number of edges is bound by the available DRAM, and could reach about 5 billion on the 64GB of DRAM available in the Alveo U200 card.
Our design can be easily scaled to compute multiple \gls{ppr} vectors in parallel, if the end-user can provide an upper bound over the number of vertices in its graphs. 
In our experiments, optimal performance results are achieved if the number of vertices does not exceed 1 million (which is still larger than what is found in many real applications), and 8 to 16 \textit{personalization} vertices are computed in parallel, using the same hardware resources required for a larger graph that does not consider multiple \gls{ppr} vertices.

The size of local memory buffers is not a limitation on the size of the graphs: first, our \gls{ppr} implementation targets graphs encountered in social network communities and e-commerce platforms, whose size does not fill the available \gls{fpga} hardware resources \cite{snapnets}; second, there exist partitioning techniques \cite{umuroglu2015vector, shan2010fpga} that handles large web-scale graphs.
Scalability to web-scale graphs, although not required in our use-case or to validate the performance of our \gls{spmv} implementation, is very interesting;
these approaches, however, are mostly orthogonal to our design and integrating them  would not demand a deep rethinking of our architecture.

\subsection{Host Integration}\label{sec:host}
Our architecture follows a host-accelerator model in which the \textit{host} (a server) communicates with the \textit{accelerator} (an \gls{fpga}) over PCIe.
Pre-processing (e.g. loading the graph) is done once at the start and not for each computation of \gls{ppr}, and it takes a negligible amount of time ($< 1\%$ of the execution time). 
Re-synthesizing the architecture is required to change the fixed-point precision, $\kappa$ or the maximum number of vertices in URAM, but not for different input graphs.
\section{Experimental Evaluation}\label{sec:experimental_results}

Our architecture is implemented on a Xilinx Alveo U200  Accelerator Card with 64 GB of DRAM (77 GB/s of total bandwidth) and equipped with a \texttt{xcu200-fsgd2104-2-e} \gls{fpga} offering 960 UltraRAM blocks of 288Kb (with 72 bits port width) and 4320 BRAM blocks with 18Kb size each. This \gls{fpga} platform is mounted on a server with an Intel Core i7-4770 CPU @ 3.40GHz with 4 cores (8 threads) and 16 GB of DRAM.
We compare our \gls{ppr} implementation against the floating-point implementation in PGX 19.3.1 \footnote{\url{docs.oracle.com/cd/E56133_01/latest/index.html}}, a powerful toolkit for in-memory graph analytics.
Its state-of-the-art implementation of \gls{ppr} \cite{hong2012green} is fully multi-threaded. Experiments with PGX were conducted on a machine equipped with two Intel Xeon E5-2680 v2 @ 2.80GHz with 10 cores (20 threads) each, and 384 GB of DRAM.
We analyze 5 versions of our architecture: 26 bits unsigned fixed-point (\texttt{Q1.25}), 24 bits (\texttt{Q1.23}), 22 bits (\texttt{Q1.21}), 20 bits (\texttt{Q1.19}), and a 32-bit floating point version (\texttt{F32}).
Lower bit-width negatively impacts the quality of results, while higher precision provides minimal gain (\cref{sec:accuracy_discussion}).
The CPU baseline uses 32 bits floating-point arithmetic, and our CPU does not support arbitrary precision. Simulated fixed-precision arithmetic resulted in lower CPU performance, and is not a meaningful comparison. 
Manually batching multiple requests in PGX through vector properties did not provide a speedup over the fast default implementation of \gls{ppr}, which is already fully exploiting the CPU \cite{zhang2018graphit}.

\renewcommand\theadalign{tl}
\renewcommand\theadfont{\bfseries}

\begin{table}
\centering
\ra{1.0}
    \caption{Summary of graph datasets used in the evaluation}
    \resizebox{1\columnwidth}{!}{
	\begin{tabular}{@{}llll@{}}
		\toprule
		\thead{Graph Distribution} & \thead{$\mathbf{|V|}$} & \thead{$\mathbf{|E|}$} & \thead{Sparsity}\\
		\midrule
		\multirow{2}{*}{\textbf{$G_{n,p}$ (Erd\H{o}s-Renyi)}} & $10^5$ & $1002178$ & $10^{-4}$ \\
		 & $2 \cdot 10^5$ & $1999249$  & $4.9 \cdot 10^{-5}$\\
		 \midrule
		\multirow{2}{*}{\textbf{Watts–Strogatz small-world}} & $10^5$ & $1000000$ & $10^{-4}$  \\
		 & $2 \cdot 10^5$ & $2000000$ & $5 \cdot 10^{-5}$ \\
		 \midrule
		\multirow{2}{*}{\textbf{Holme and Kim powerlaw}} & $10^5$ & $999845$ & $0.99 \cdot 10^{-4}$ \\
		 & $2 \cdot 10^5$ & $1999825$ & $4.9 \cdot 10^{-5}$ \\
	 	\midrule
		\textbf{Amazon co-purchasing network} & $128000$ & $443378$  & $2.7 \cdot 10^{-5}$ \\
		\textbf{Twitter social circles} & $81306$ & $1572670$  & $2.3 \cdot 10^{-4}$ \\
		\bottomrule
	\end{tabular}
    }
    \label{tab:graphs}
\end{table}

\renewcommand\theadalign{tl}
\renewcommand\theadfont{\bfseries}
\setlength\tabcolsep{3pt}

\begin{table}
\centering
\ra{1.0}
    \caption{Resource usage, power consumption of our design.
    Other bit-widths, omitted for brevity, show the same trends}
    \resizebox{1\linewidth}{!}{
	\begin{tabular}{@{}llllllll@{}}
		\toprule
	\thead{Bit-width} & \thead{BRAM} & \thead{DSP} & \thead{FF} & \thead{LUT}  & \thead{URAM} & \thead{Clock\\(MHz)} & \thead{Power\\Cons.}\\
	   
	    \midrule
    	
    	
		20 bits & 14\% & 3\% & 4\% & 26\% & 20\% & 220 & 34 W \\
		 
		26 bits & 14\% & 3\% & 4\% & 38\% & 20\% & 200 & 35 W \\
		
		32 bits, \textbf{float} & 14\% & 48\% & 35\% & 89\% & 26\% & 115 & 40 W \\
		

		
		 
		 
		
        \midrule
        \textbf{Available} & 4320 & 6840 & 2364480 & 1182240 & 960 & & \\
		\bottomrule
	\end{tabular}
    }
    \label{tab:resources}
\end{table}

Our experimental setup contains 8 graphs (\cref{tab:graphs}): 6 are generated using different statistical distributions offered by the Python \texttt{networkx} library\footnote{\url{networkx.github.io/documentation/stable/}}, while 2 are real-world graphs from the Stanford Large Network Dataset Collection \cite{snapnets}.
Synthetic graphs are consistent in size, edge distribution, and sparsity to real-world graphs used in e-commerce and social network communities \cite{snapnets}; their \gls{coo} representation has size in line with recent work on sparse matrices on \glspl{fpga} \cite{grigoras2018instance}.
Synthetic graphs with identical sizes highlight how trends are similar across distributions (\cref{sec:time}, \cref{sec:accuracy}), and we can extract insights on the convergence and precision of \gls{ppr} as we change input graph and bit-width.

\subsection{Execution time}\label{sec:time}

We measure for each graph the time required to compute the \gls{ppr} values for 100 random \textit{personalization} vertices, to simulate a realistic batch workload performed by social networks and e-commerce platforms. 
Time spent transferring results from \gls{fpga} to CPU is included, and is negligible compared to the total execution time. All tests are executed with an $\alpha$ of $0.85$, for $10$ iterations each (even a low amount of iterations is enough for convergence, see \cref{sec:accuracy_discussion}).

\begin{figure}
    \centering
    \includegraphics[width=0.475\textwidth]{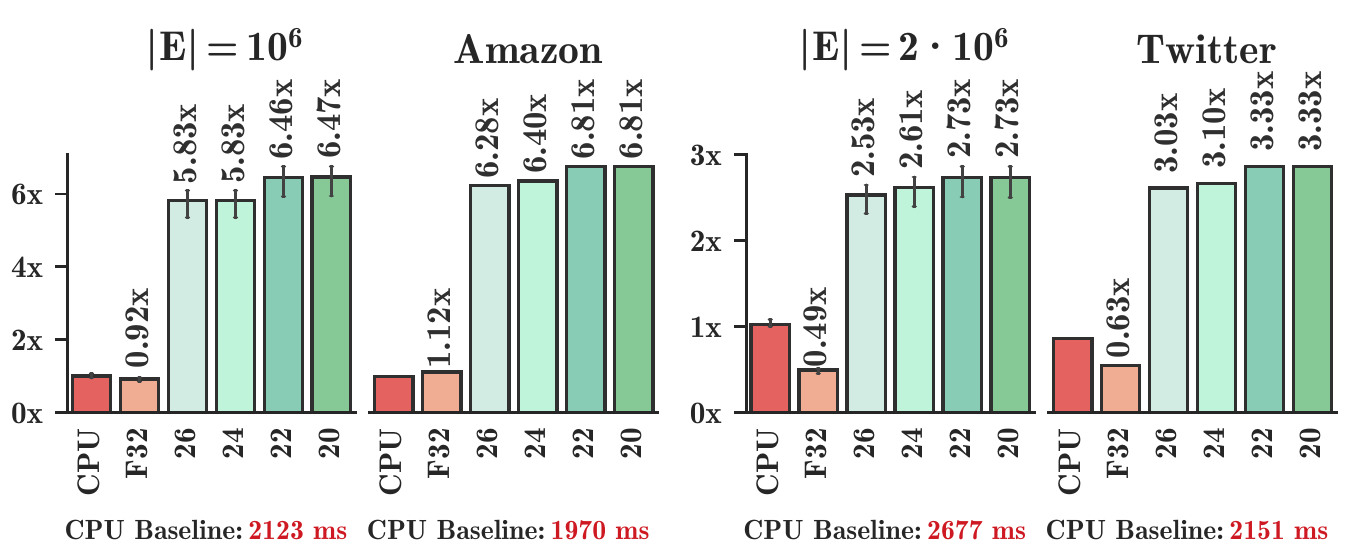}\\
    \caption{
    Speedup of our \gls{fpga} implementation (y-axis) w.r.t. the CPU baseline, for decreasing bit-widths (x-axis).
    }
    \label{fig:exec_times}
\end{figure}

\Cref{fig:exec_times} reports the speedups of different fixed-point sizes compared to the CPU baseline and to an equivalent 32-bits floating-point FPGA architecture. Reducing bit-width
shows a positive correlation with clock speed, and higher speedups. On graphs with around $10^6$ edges we obtain up to \textbf{6.47x} speedup, thanks to the reduced bit-width and the ability to compute 8 \gls{ppr} vectors at once. Results for synthetic graphs are averaged, as no difference was observed among distributions. We achieve similar results on real-world graphs, with up to \textbf{6.8x} speedup on the highly sparse Amazon co-purchasing network. 
The time required by the \gls{fpga} for 100 random requests ranges from 280 ms for Amazon to 1000 ms for larger graphs, which is in line with the real-time requirement of our use-case.
The floating-point FPGA architecture is 6 times slower than the fixed-point designs, with larger DSP usage (48\% vs 3\%), and negligible accuracy gain compared to 26-bits fixed-point  (\cref{fig:errors}). 

The clock frequency is between 200 and 220 MHz, but we can reach up to 350 MHz with lower number of concurrent \gls{ppr} vertices $\kappa$. The clock speed increases sublinearly w.r.t $\kappa$ above 200 MHz, limiting the benefits of very low $\kappa$. On larger graphs the speedups are less significant, as higher URAM utilization negatively impacts the clock frequency due to routing congestion. In our experiments, doubling the size of the \gls{ppr} buffers lowers the clock speed by around 35-40\%.
Resources utilization (summarized in \cref{tab:resources} for $\kappa=8$), is minimal for BRAM, DSPs and registers and is not impacted by fixed-point bit-width and \gls{ppr} vector size. URAM usage grows linearly with \gls{ppr} vector size (from 20\% to 40\% in our experiments).


\subsection{Energy Efficiency}

Our FPGA architecture uses 35W during execution, and increasing the \gls{ppr} buffer or the fixed-point bit-width does not seem to affect the power consumption.
The CPUs consume around 230W, and our architecture provides a Performance/Watt gain from \textbf{16.5x} to \textbf{42x} compared to it (geomean \textbf{28.2x}). Even against a faster CPU or a GPU, our architecture is likely to offer higher energy efficiency. Using fixed-point provides \textbf{5x} higher energy efficiency over the equivalent floating-point design, which however provides \textbf{2.5x}-\textbf{5x} higher energy efficiency than the CPU baseline (geomean \textbf{4.3x}).

\subsection{Accuracy Analysis}\label{sec:accuracy}

\begin{figure}
    \centering
    \includegraphics[width=0.46\textwidth]{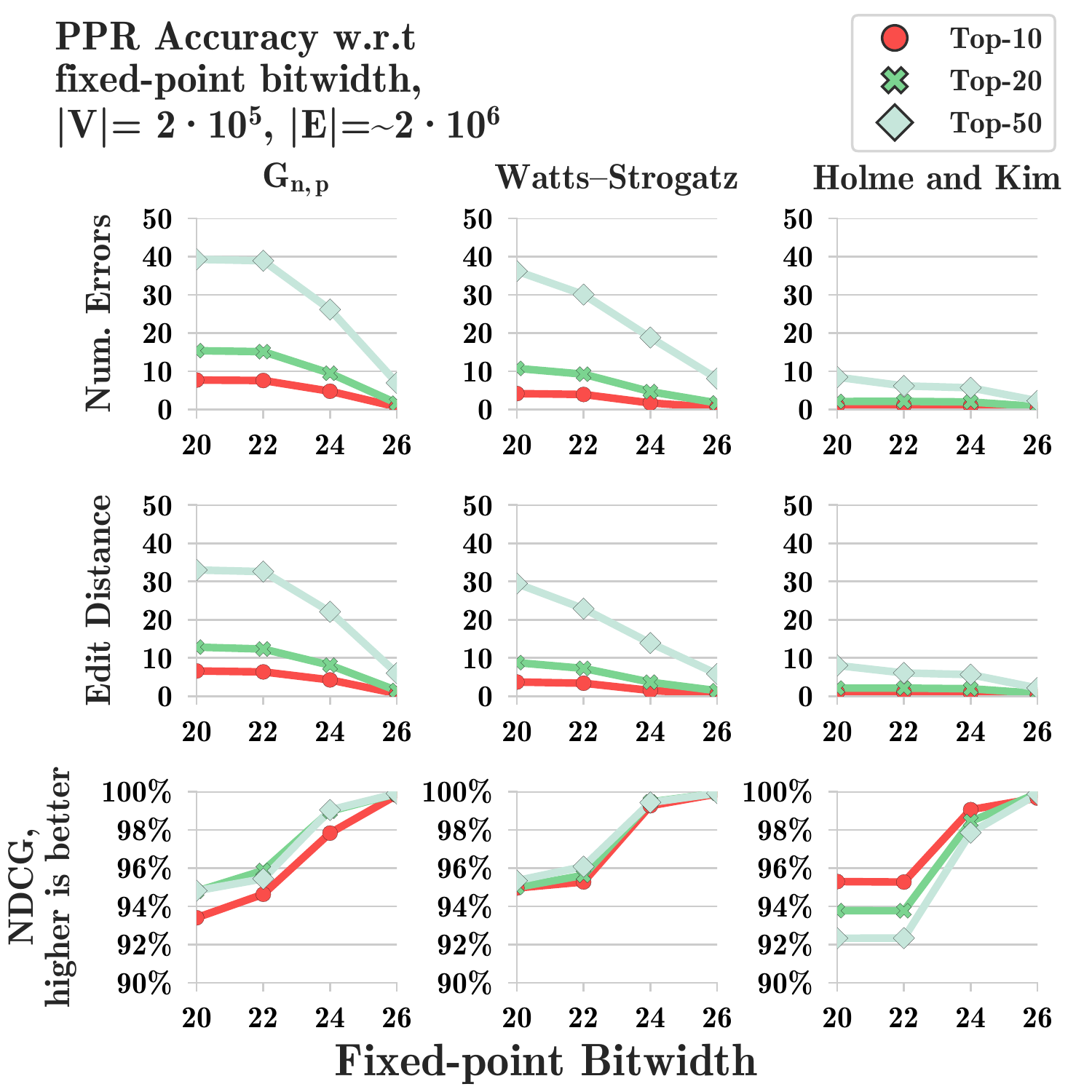}
    \caption{Accuracy metrics for graphs with $\mathbf{2 \cdot 10^6}$ edges, with increasing fixed-point bit-width. \textbf{Number of errors} and \textbf{edit distance} should be low, while \textbf{NDCG} must be close to 100\%.
    }
    \label{fig:errors}
\end{figure}

We compared the accuracy of the rankings obtained with fixed-point precision (after 10 iterations of \gls{ppr}) with the ones of the CPU implementation at convergence (with at least 100 iterations), using common \gls{ir} ranking metrics \cite{said2015replicable}. 100 iterations are enough to reach convergence even in web-scale graphs \cite{langville2004deeper}, although 10 iterations would often suffice (\cref{fig:errors}, \cref{fig:convergence}).

\subsubsection{Accuracy metrics}\label{sec:metrics}

First, we look at the \textbf{number of errors}, i.e. the number of vertices with wrong ranking in the top 10, 20 and 50 compared to the CPU.
This metric is very coarse-grained, as a single mistake can greatly affect the ranking:
for example, if the correct top-4 values are $\{2,4,8,6\}$ and we retrieve $\{4,8,6,2\}$, this  metric reports 4 errors, although only a single value is displaced.


\textbf{Edit Distance} counts how many operations are needed to transform one sequence of top-N vertices into another \cite{levenshtein1966binary}; it handles ordering shifts: in the previous example the edit distance is just 1, as we insert $2$ at the beginning and ignore values after the first N.


\textbf{\gls{ndcg}} \cite{jarvelin2002cumulated} is commonly used to evaluate recommender systems: it dampens the \textit{relevance} of a vertex by a logarithmic factor such that
highly ranked vertices contribute more to the \textit{cumulative gain}.
Given a vector of \gls{ppr} scores, $rel_i = |V| - i$ is the \textit{relevance} of the $i$-th vertex, and we define \gls{dcg} as in \cref{eq:dcg}. \gls{dcg} is normalized by the \textit{Ideal DCG} of the CPU implementation.

\begin{equation*}
\refstepcounter{equation}\latexlabel{eq:dcg}
\refstepcounter{equation}\latexlabel{eq:ndcg}
\mathbf{DCG} = \sum_{i = 1}^{|V|} \frac{rel_i}{log_2(i + 1)}\qquad \mathbf{nDCG} = \frac{DCG}{IDCG}
\tag{\ref{eq:dcg}}
\end{equation*}

\subsubsection{Accuracy Discussion}\label{sec:accuracy_discussion}

\begin{figure}[htp]
  \includegraphics[width=0.465\textwidth]{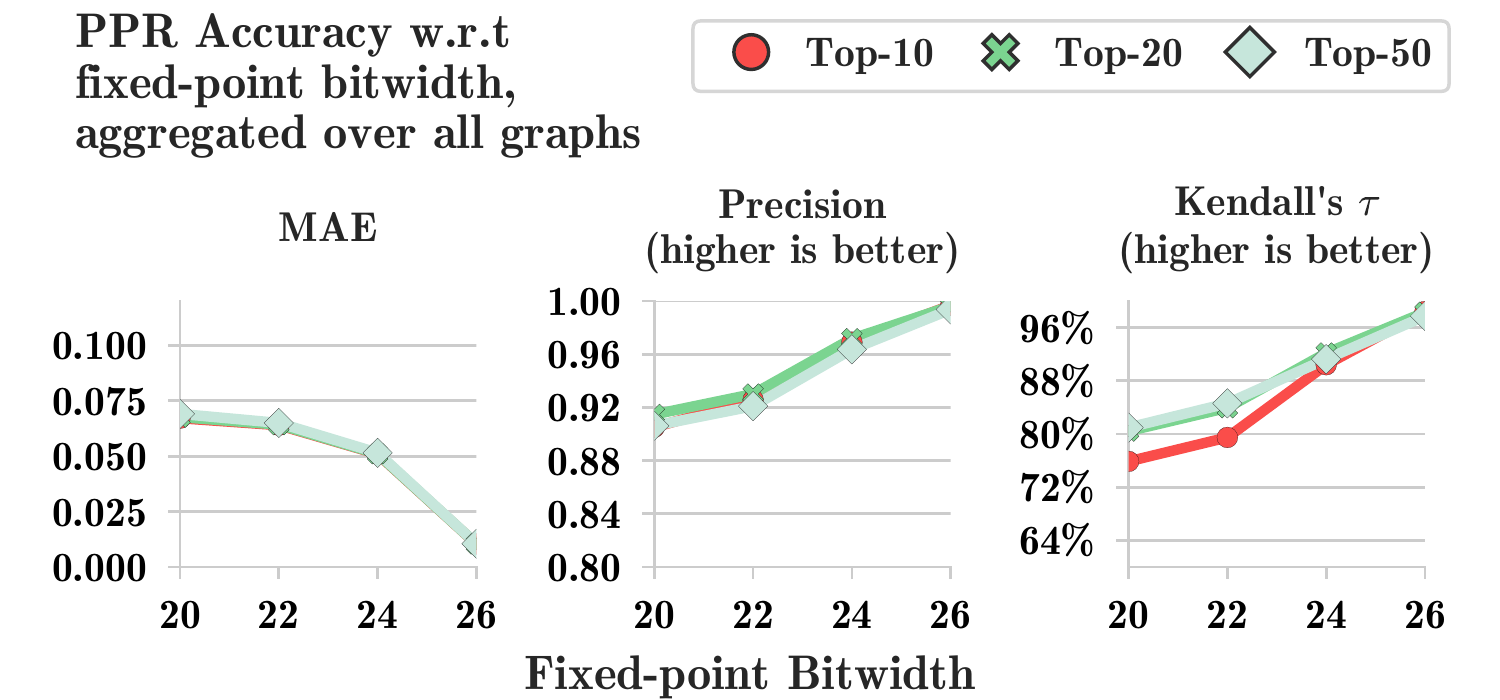}
    \caption{Aggregated accuracy metrics show trends in-line with \cref{fig:errors}, and even low bit-width provides good predictions}
    \label{fig:errors_agg_a}
\end{figure}

\Cref{fig:errors} shows how metrics change by lowering the fixed-point bit-width, for each of the $2 \cdot 10^6$ edges graphs. 
\Cref{fig:errors_agg_a} shows additional accuracy metrics, aggregated on all graphs:  \textit{Mean Average Error} (MAE), \textit{Precision} and \textit{Kendall's} $\tau$. MAE evaluates how far \gls{fpga} \gls{ppr} values are from the correct ones, while Precision measures the top-N correctness without looking at the vertices order; just 20 bits are enough to retrieve 90\% of the best top-50 items. Kendall's $\tau$ is a ranking metric that penalizes out-of-order predictions \cite{shani2011evaluating}.
Results in \cref{fig:errors_agg_a} are similar to \cref{fig:errors}, with MAE and Precision mostly unaffected by a larger set of predictions.

\begin{figure}[htp]
  \includegraphics[width=0.465\textwidth]{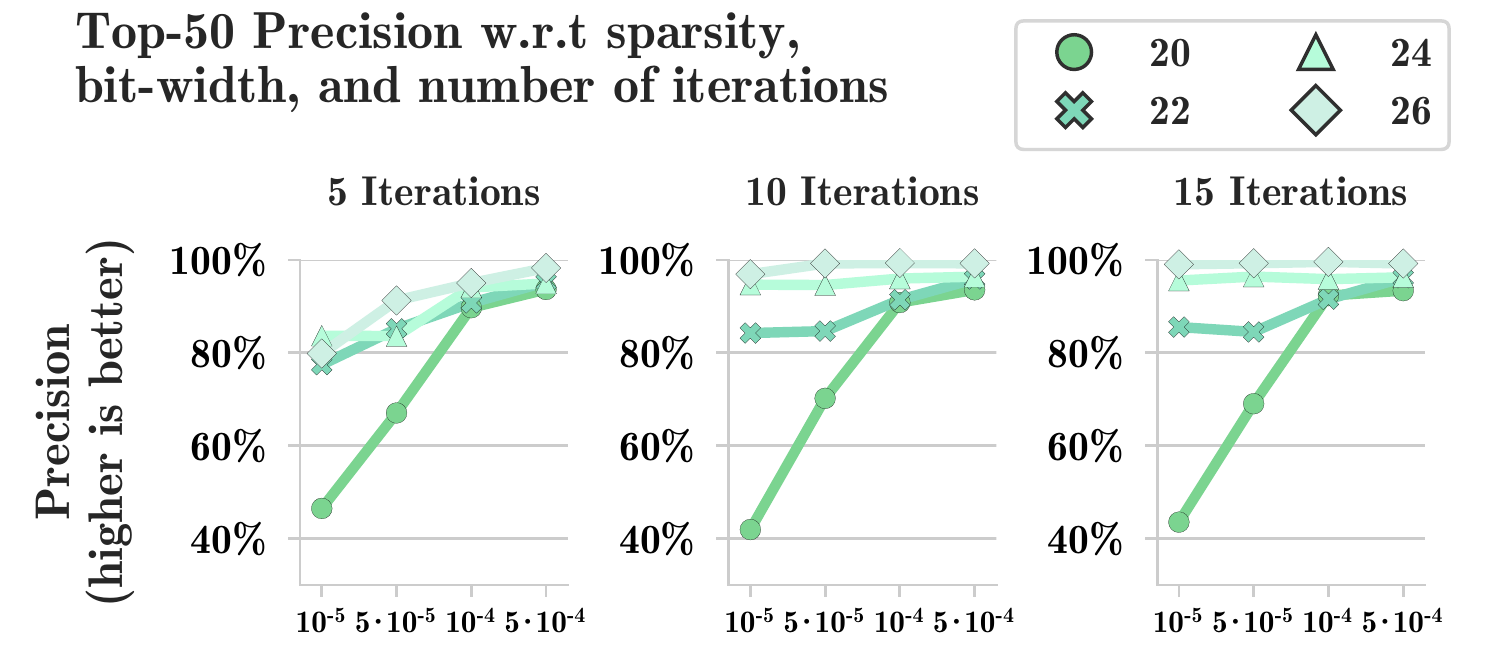}
    \caption{Sparsity does not affect accuracy, except for very low bit-width, and 10 iterations are enough for convergence. Other metrics show similar trends as the top-50 Precision}
    \label{fig:errors_agg_b}
\end{figure}

Increasing bit-width is always beneficial, with diminishing returns. Using 26 bits provides near-to-perfect results, although even 22 or 24 bits provide satisfactory results, with more than half of the vertices being ranked correctly. 22 bits show a top-10 edit distance of 3 and an NDCG value $>$ 95\%. With 26 bits, the top-20 edit distance is $<$ 3, i.e. only 3 values in the first 20 are out-of-place. Results are impacted by graph distribution: \textit{Holme and Kim} graphs, for which errors are lower, have dense communities, similarly to real social networks, while the behavior of the other 2 models is more unpredictable.
Sparsity has a minor impact on accuracy (\cref{fig:errors_agg_b}): very low bit-width suffers from high sparsity, but in general results are consistent with \Cref{fig:errors}. 
We display the top-50 precision due to space limitations, but other metrics show identical behaviors. 

\begin{figure}
    \centering
    \includegraphics[width=0.475\textwidth]{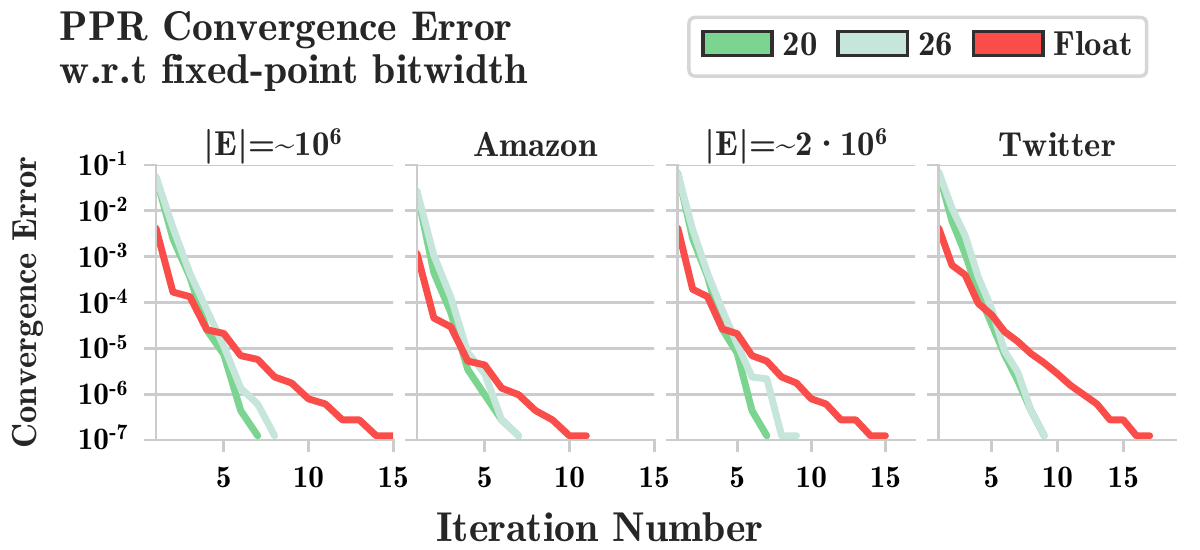}
    \caption{Fixed-point gives 2x faster convergence compared to floating-point. Lines are truncated for error below $\mathbf{10^{-7}}$}
    \label{fig:convergence}
\end{figure}

Fixed-point arithmetic produces faster convergence (\cref{fig:convergence}). We measure, after each iteration, the Euclidean norm of new and previous \gls{ppr} values, to evaluate convergence. Less than 20 iterations are always enough for convergence, and even 10 iterations provide an error below $10^{-6}$ (a common convergence threshold for \gls{ppr} \cite{nvgraph}). Fixed-point arithmetic converges twice as fast compared to floating-point, while preserving accuracy (\cref{fig:errors}). In real computations, \gls{ppr} stops when the error is below a threshold: a 2x faster convergence immediately translates to an additional 2x speedup over a floating-point implementation.
Lower bit-width provides 10-20\% faster convergence in synthetic graphs.

\section{Conclusion and future work}
We presented a high-performance \gls{fpga} implementation of a \gls{coo} \gls{spmv} algorithm that leverages data-flow computation and reduced-precision fixed-point arithmetic. 
We have shown how our architecture accelerates the \gls{ppr} algorithm and outperform a state-of-the-art CPU implementation by up to \textbf{6.8x}, with up to \textbf{42x} higher energy efficiency.
With just 26-bits fixed-point values we guarantee a speedup above 5.8x with negligible accuracy loss, with \textbf{2x} faster convergence: average top-10 edit is distance below 1 and \gls{ndcg} is above 99.9\% compared to the CPU, showing how graph ranking algorithms can benefit from approximate computing.


Although the present work focuses on the design of a fixed point \gls{coo} \gls{spmv} for a specific use-case and is not a general-purpose graph engine, we deem valuable to integrate partitioning techniques \cite{umuroglu2015vector, shan2010fpga} and support web-scale graphs, and study the optimal trade-off between partitioning overheads and \gls{fpga} resource utilization.
A comparison against modern GPUs is also very interesting: we omitted detailed GPU analyses as we currently lack high-end GPUs comparable to the Alveo U200. The GTX960 at our disposal is as fast as the CPU baseline using nvGRAPH, although Nvidia claims a 3x speedup using a faster Tesla M40 against a CPU similar to ours \cite{nvgraph}. 
We will also apply our reduced precision \gls{spmv} on other use-cases, such as graph embeddings \cite{geng2019uwb}.

\bibliographystyle{ACM-Reference-Format}
\bibliography{references}

\end{document}